\documentclass[a4paper,10pt]{article}

\begin{document}

\title{Inhomogeneities in the Universe and the Fitting Problem}

\author{{\bf Marie-No\"elle C\'el\'erier}
\\
Laboratoire Univers et Th\'eories (LUTH), \\ 
Observatoire de Paris-Meudon, \\
5 place Jules Janssen, 92195 Meudon cedex, France \\
Email: marie-noelle.celerier@obspm.fr}

\maketitle

\begin{abstract}

Observational cosmology provides us with a large number of high precision data which are used to derive models trying to reproduce ``on the mean'' our observable patch of the Universe. Most of these attempts are achieved in the framework of a Friedmann-Lema\^itre cosmology where large scale homogeneity is assumed, following the so-called Cosmological ``Principle''. However, we know, from the observation of structures at increasing scales (e. g., more than 400 Mpc for the lately found Sloan Great Wall), that these models are only approximations of a smoothed or averaged inhomogeneous underlying patern. Anyhow, when modelling the Universe, the usual method is to use continuous functions representing, e. g., energy-density, pressure, or other kinematical scalars of the velocity field, implicitly assuming that they represent volume averages of the corresponding fine-scale inhomogeneous quantities, then put them into the Einstein equations which are solved to give the model and its dependance upon a number of parameters arbitrarily defined. In General Relativity, such a method is very much involved since the equations which determine the metric tensor and the quantities calculated from it are highly nonlinear. The main issue is the non-commutating property of the two operations: averaging the metric and calculating the Einstein tensor. The question raised by the method consisting of determining the parameters of an a priori assumed FLRW model from observational data is the ``fitting problem'' brought to general attention by Ellis and Stoeger in the 80's. This problem has recently experienced a reniewed attention due to the amount of available data and the increase of the minimum scale at which homogeneity can be assumed. We propose a discussion of this issue in the light of the latest developments of observational and theoretical cosmology.

\end{abstract}

\maketitle

\section{Introduction}

Cosmology is aimed at finding the Universe model best fitting the data obtained from observations.

In the standard approach, one makes some a priori assumption about the Universe geometry based on philosophical grounds. The philosophical assumption generally retained is the Copernican principle which states that we are not located at a special place in the Universe and that what we observe must be the same as what would be observed from any other place. Combined with the observation of  the Cosmic Microwave Background Radiation (CMBR) quasi-isotropy, it implies that the Universe should be homogeneous, at least on some large, but undetermined, scales. This is called the cosmological principle. The thus assumed homogeneity leads most cosmologists to use the Friedmann-Lema\^itre-Robertson-Walker (FLRW) model to represent the Universe. Their objective is therefore to determine the free parameters of such a model.

However, we know, from the observation of structures at increasing scales (e. g., more than 400 Mpc for the lately found Sloan Great Wall \cite{JG05}), that these models are only approximations of a smoothed or averaged inhomogeneous underlying patern.

An alternative would be to start from the observations and determine the space-time geometry without a priori assumptions.

However, these approaches are both unsatisfactory. The standard approach, because the Universe is obviously not FLRW, at least on some (undetermined) scales. Therefore it does not provide any clue as to on what scale this model is applicable and it does not consider relations between descriptions of the Universe at different inhomogeneity scales. The alternative, because of its implementation practical difficulties, since it is an impossible task to construct a model reproducing all the structures down to the smallest scales.

We are therefore faced with the necessity to find another way to deal with this problem.

\section{Averaging}

\subsection{The issue}

Since our Universe is observed to be inhomogeneous at the scales of galaxies, clusters, super-clusters, etc. the cosmological issue involve always explicit or implicit averaging processes to smooth out the underlying inhomogeneities. However, the explicit implementation of such processes is confronted to a number of difficulties issued from the nature of the General Relativity theory (GR).
  
In the cosmological context, the gravitational field is created by a point-like discrete matter distribution, i. e., the galaxies. A problem therefore arises when we want to apply GR to quantities such as matter distribution, pressure, or other kinematical scalars of the velocity field, implicitly assuming that they represent volume averages of the corresponding fine-scale quantities which we consider as smooth and which we represent by continuous functions.
 
Moreover, the scale of averaging, when using FLRW models, has never been explicitly agreed upon while the result of averaging obviously depends on it.

Another pitfall is that covariant volume averages are well-defined quantities for scalars only. They are difficult to define for vectors, and all the more for tensors. This is the reason why most of the averaging schemes available on the market only consider the volume averages of scalar quantities.

However, volume averages are tricky since the results of the required splitting of a generic curved manifold into time and 3-space depend on the choice of the hyper-surfaces on which the average is performed. Different choices leading to different results, such volume averages must be used with care and precisely defined.

Gauge problems arise also since scalar quantities only are invariant under coordinate transformations, not tensors.

But the main issue is that the GR equations are non-linear which implies non-commutation of averaging the metric with calculating the Einstein tensor. Therefore, averaging does modify Einstein's equations by an effective energy term, even for FLRW backgrounds.
  
The consequences are: (i) that a lumpy Universe of which the geometry is FLRW on average evolves differently, i. e., not according to the Friedmann equation; (ii) that the averaged Einstein equations are not the same when applied on different scales; (iii) that expansion affects local measurements within expanding regions (voids) differently from those within a (non expanding) bound system and therefore, e. g., CMBR temperature measured at the volume average in voids is lower than for bound system observers which would imply to recalibrate the cosmological parameters associated with the radiation-dominated era \cite{DW07}. 
  
\subsection{Main proposed averaging schemes}

In this section, the main averaging schemes which have been applied to the cosmological issue are set out. Many other proposals have been made, which can be found in the literature, but our purpose here is not to be exhaustive and we refer the interested reader to Ref. \cite{AK97}.
  
The best systematic and coherent attempt to date to set up the problem generically and to propose a covariant spacetime averaging procedure is Macroscopic Gravity developed by Zalaletdinov \cite{RZ92}. In this scheme, the Einstein equations are modified by gravitational correction terms which take the form of a correlation tensor.

Buchert and Carfora \cite{BC00} have derived a covariant averaging process well-defined for a given foliation of space-time and for applications to scalar quantities. They make a 3+1 splitting of space-time and average on the space-like hyper-surfaces. The backreactions are computed from a non-perturbative averaging of scalar quantities. ``Bare'' and ``dressed'' parameters are defined. The former provides the framework for interpreting the observations with a ``Friedmann bias'', i. e., as if the observer was living in a Friedmannian Universe. It is the ``standard'' approach. The latter represents the actual inhomogeneous cosmological model, spatially averaged.

Other authors have proposed averaging within approximation schemes. The main one to have been used in the cosmological context is the proposal by Futamase \cite{TF88}. The author begins with a 3+1 splitting of space-time, then he averages on the spatial hyper-surfaces thus defined.  The backreactions are perturbatively computed.

\section{The fitting problem}

The basic idea first put forward by Ellis \cite{GE84} then developed by Ellis and Stoeger \cite{ES87} is that it is not clear that merely averaging the observed data is an appropriate procedure. Therefore they do not a priori assume the Universe is well-described by a FLRW model at all scales and times but nevertheless decide to use it for practical reasons.

The issue is as follows. They contemplate a lumpy realistic cosmological model including all inhomogeneities down to some length scale $l$ and an idealized smoothed out FLRW model at some length scale $L>l$.
 
The problem, as they formulated it, is to determine at which smoothing scale $L$ a best-fit can be obtained between the geometry of these two models.

\subsection{Adequate fitting procedure}
  
These authors claim that an adequate fitting procedure must have the following properties.
 
It must be based on observational criteria, using all the relevant available data in a coherent way.

It must not only determine best-fit parameters for the FLRW model of Universe but also give a space-time fit between this FLRW model and the lumpy Universe and characterize the goodness-of-fit achieved while specifying criteria of acceptable fit. We would therefore reject {\it any} FLRW model for the observed lumpy Universe which might not fulfill these criteria.

It must define a relevant averaging process to pass from the lumpy Universe to its representation by a FLRW model.

And last but not least, it must determine the scale dependence of this process and in particular: (i) a scale of uniformity, i. e., a scale beyond which an average of the lumpy Universe gives a FLRW model to a good approximation, (ii) and a length scale above which the Hubble law apply to local physical system, to determine the minimum scale of expanding unbound systems.
 
\subsection{Volume Matching, the main steps}

The Volume Matching procedure proposed by Ellis and Stoeger \cite{ES87} is a null data fitting related directly to astronomical observations realized on our past light cone. We only give here a summary of the method and refer the interested reader to the original article. Its main steps are:
  
  \begin{itemize} 
  
\item Choose a correspondence between the vertex points (the observers) and the 4-velocities at these points for both past light cones in the lumpy and FLRW models.

\item Isotropy on the past light cones to obtain averaged (spherically symmetric) geometries.

\item Compare these geometries down the two light cones and obtain a best-fit FLRW model and its degree of approximation of the lumpy Universe on the past light cone.

\item Construct the best-fit space-time inside and outside the null cone and estimate how good the fit is off the light cone.
  
  \end{itemize}

\section{Implementation examples}

All the above described procedures have been implemented in the cosmological framework to estimate the errors made in applying the ``standard'' method of assuming an all scale homogeneous Universe to describe an underlying lumpy one.
 
For instance, Hellaby \cite{CH88} has computed the error made when using averaging procedures compared to the Volume Matching of FLRW models to inhomogeneous Lema\^itre-Tolman-Bondi (LTB) solutions with realistic density profiles. He has found that the mean density and pressure of the averaged FLRW models are 10-30\% underestimated as regards the volume matched ones.

Coley and Pelavas \cite{CP07} have studied the outcomes of Macroscopic Gravity. They have found that the correlation tensor in a FLRW background exhibits the form of a spatial curvature. For a non-FLRW background this correlation tensor mimics spatial curvature to which an effective anisotropic fluid term is added.

Buchert's averaging scheme has been used in a number of works.

Two effects that quantify the difference between ``bare'' and `dressed'' parameters have been identified in this framework: the ``curvature backreaction'' and the ``volume effect''. The regional curvature backreaction is built from scalar invariants of the intrinsic curvature. It features two positive-definite parts, the scalar curvature amplitude fluctuations and fluctuations in metrical anisotropies. Depending on which part dominates, one obtains an under or overestimate of the actual averaged scalar curvature. The volume effect is due to the difference between the volume of the smoothed region and the actual volume of the lumpy region. Buchert and Carfora \cite{BC03} have estimated the volume effect alone on the mean density in a naive swiss-cheese model. They obtain a 64\% increase of the $\Omega_M$ parameter, which, in a flat FLRW model, implies reducing dark energy from $\Omega_{\Lambda}\approx 0.7$ to $\Omega_{\Lambda} \approx 0.5$. However, other effects, such as light-cone effects, remain to be actually taken into account in this procedure.

Curvature dominated unbound LTB models averaged with Buchert's method have been shown to exhibit late time accelerating expansion \cite{PS06,JM06}. However, the effect remains to be quantified.

R\"as\"anen \cite{SR06} have also used Buchert's recipe to study a simplified model of structure formation composed of averaged FLRW under-dense and over-dense patches which leads to late time cosmological acceleration, still to be quantified.

Note that the possibility of explaining the SN Ia luminosity dimming, and its assumed Friedmannian counterpart, accelerated expansion, as an effect of the underlying structures has been reviewed in Ref. \cite{MNC07}. This review includes not only attempts to solve the averaging or fitting problems but also uses of exact inhomogeneous models to deal with the cosmological constant and coincidence problems.

The last result we will cite here has been derived by Bildhauer and Futamase \cite{BF91} with the use of Futamase's averaging scheme. Considering a simple flat cosmological model analyzed in the framework of the pancake theory for structure formation \cite{YZ70}, they obtain an increase of the Universe age from 13.3 Gyr for the ``standard'' approach to above 17 Gyr for the new one.

However, the issue of calculating corrections to Einstein's equations, in a cosmological setting and with an averaging procedure, as physically relevant quantities remains an open question partly due to possible ambiguities in the available averaging schemes. In particular, it might be argued that the added terms could be gauge artifacts.

This problem has been partly solved by Paranjape and Singh \cite{PS07} who computed, in the framework of Macroscopic Gravity, space-time scalar corrections to the Friedmannian equations, independent of the choice of coordinates in the averaged manifold, but without being able to escape the problem of dependence on the gauge choice which is inherent to the large scale homogeneity assumption. These authors explicitly construct these scalars in terms of the underlying inhomogeneous geometry and show that the formal structure of the corrections in the peculiar gauge retained is identical to that of analogous corrections derived by Buchert.

\section{Conclusion}
  
We have shown that the standard way of determining the cosmological parameters of an a priori assumed large scale homogeneous Universe implies systematic errors. Moreover, the scale above which the Universe can be validly represented by a FLRW model is generally not explicitly determined while the results of the data analysis obviously depends on it. Implementing a correct fitting procedure is therefore mandatory but challenging.

Tentative evaluations of the errors pertaining to the ``standard'' procedure give from 10\% to 64\%, but remain to be improved with more realistic models of the lumpy underlying Universe and more robust averaging recipes. However, these first evaluations show that the effect might be not negligible and must be taken into account to deal with the cosmological issue, since it will be an unavoidable step towards the development of ``precision'' cosmology.

Projects aimed at determining the Universe geometry from cosmological data while verifying and quantifying homogeneity rather than assuming it are currently underway \cite{LH07} and must be developed. This is one of the most important challenges for cosmology.

\end{document}